\documentclass[iop,apj]{emulateapj}
\usepackage[backref,breaklinks,colorlinks,citecolor=blue,allcolors=blue]{hyperref}

\usepackage{amsmath}
\usepackage{color}

\newcommand{\Tr}{{\text{Tr}}}
\newcommand{\bb}{{\boldsymbol{b}}}
\newcommand{\bB}{{\boldsymbol{B}}}
\newcommand{\bC}{{\boldsymbol{C}}}
\newcommand{\bD}{{\boldsymbol{D}}}
\newcommand{\bF}{{\boldsymbol{F}}}

\newcommand{\bG}{{\boldsymbol{G}}}
\newcommand{\bI}{{\boldsymbol{I}}}
\newcommand{\bJ}{{\boldsymbol{J}}}
\newcommand{\bK}{{\boldsymbol{K}}}
\newcommand{\mL}{{\mathcal{L}}}
\newcommand{\bmM}{{\boldsymbol{\mathcal{M}}}}

\newcommand{\mO}{{\mathcal{O}}}
\newcommand{\bP}{{\boldsymbol{P}}}
\newcommand{\bQ}{{\boldsymbol{Q}}}
\newcommand{\bS}{{\boldsymbol{S}}}
\newcommand{\bU}{{\boldsymbol{U}}}
\newcommand{\bV}{{\boldsymbol{V}}}
\newcommand{\bW}{{\boldsymbol{W}}}
\newcommand{\bx}{{\boldsymbol{x}}}
\newcommand{\by}{{\boldsymbol{y}}}
\newcommand{\bmu}{{\boldsymbol{\mu}}}
\newcommand{\bn}{{\boldsymbol{n}}}
\newcommand{\bxi}{{\boldsymbol{\xi}}}
\newcommand{\bdelta}{{\boldsymbol{\delta}}}
\newcommand{\bDelta}{{\boldsymbol{\Delta}}}
\newcommand{\bgamma}{{\boldsymbol{\gamma}}}
\newcommand{\bGamma}{{\boldsymbol{\Gamma}}}
\newcommand{\bPhi}{{\boldsymbol{\Phi}}}
\newcommand{\bPsi}{{\boldsymbol{\Psi}}}
\newcommand{\bLambda}{{\boldsymbol{\Lambda}}}
\newcommand{\bSigma}{{\boldsymbol{\Sigma}}}
\newcommand{\bPi}{{\boldsymbol{\Pi}}}

\newcommand{\bZ}{{\boldsymbol{Z}}}
\newcommand{\bzero}{{\boldsymbol{0}}}

\shorttitle{Global 21-cm signal extraction I: Separation framework}
\shortauthors{Tauscher et al.}

\begin{document}

\title{Global 21-cm signal extraction from foreground and instrumental effects I: Pattern recognition framework for separation using training sets}

\author{Keith Tauscher\altaffilmark{1,2}, David Rapetti\altaffilmark{1,3}, Jack~O.~Burns\altaffilmark{1}, Eric Switzer\altaffilmark{4}}
\affil{$^{1}${Center for Astrophysics and Space Astronomy, Department of Astrophysical and Planetary Science, University of Colorado, Boulder, CO 80309, USA}}
\affil{$^{2}${Department of Physics, University of Colorado, Boulder, CO 80309, USA}}
\affil{$^{3}${NASA Ames Research Center, Moffett Field, CA 94035, USA}}
\affil{$^{4}${NASA Goddard Space Flight Center, Greenbelt, MD 20771, USA}}

\email{Keith.Tauscher@colorado.edu}

\begin{abstract}
The sky-averaged (global) highly redshifted 21-cm spectrum from neutral hydrogen is expected to appear in the VHF range of $\sim20-200$ MHz and its spectral shape and strength are determined by the heating properties of the first stars and black holes, by the nature and duration of reionization, and by the presence or absence of exotic physics. Measurements of the global signal would therefore provide us with a wealth of astrophysical and cosmological knowledge. However, the signal has not yet been detected because it must be seen through strong foregrounds weighted by a large beam, instrumental calibration errors, and ionospheric, ground and radio-frequency-interference effects, which we collectively refer to as ``systematics''. Here, we present a signal extraction method for global signal experiments which uses Singular Value Decomposition (SVD) of ``training sets'' to produce systematics basis functions specifically suited to each observation. Instead of requiring precise absolute knowledge of the systematics, our method effectively requires precise knowledge of how the systematics can vary. After calculating eigenmodes for the signal and systematics, we perform a weighted least square fit of the corresponding coefficients and select the number of modes to include by minimizing an information criterion. We compare the performance of the signal extraction when minimizing various information criteria and find that minimizing the Deviance Information Criterion (DIC) most consistently yields unbiased fits. The methods used here are built into our widely applicable, publicly available Python package, \texttt{pylinex}, which analytically calculates constraints on signals and systematics from given data, errors, and training sets.
\end{abstract}


\section{Introduction}

The highly redshifted 21-cm spectrum from neutral hydrogen in the early Universe has become an important observational target in the astrophysics community. This is because it fills outstanding gaps in our knowledge of the Universe between the formation of the first neutral atoms $\sim 4\times 10^5$ years after the Big Bang (recombination) and when the earliest stars seen to date by the Hubble Space Telescope existed about $\sim 10^9$ years later. One of these knowledge gaps concerns the heating of the gas in the InterGalactic Medium (IGM) around the time when the first stars formed and the first black holes began accretion. This heating changes the spin temperature of the IGM gas, changing how strongly it emits the 21-cm line \citep{Madau:1997,Mirocha:2017a,Cohen:2017}. Since the 21-cm line is a quantum transition of neutral hydrogen \citep{Ewen:1951}, its brightness temperature is modulated by the neutral fraction of the IGM gas. The other knowledge gap addressable by the 21-cm signal is therefore the nature of the reionization of the Universe's hydrogen, including both when it started and how it progressed. \cite{Furlanetto:2006}, \cite{Loeb:2013}, \cite{Morales:2010}, and \cite{Pritchard:2012} provide in depth studies of 21-cm cosmology.

Experiments made to measure the highly redshifted 21-cm line come in two different varieties---power spectrum and global signal---with different science goals. The power spectrum reflects spatial anisotropies in 21-cm emission whereas the global signal quantifies its isotropic component. For both cases, but particularly for the former, physical parameter inference is difficult because the posterior distributions are broad and the likelihood functions needed to numerically explore them often require a prohibitively long time to calculate for a Markov Chain Monte Carlo (MCMC) exploration. Recently, to avoid this challenge, \cite{Kern:2017} proposed an emulation technique which uses a Gaussian Process model to drastically speed up the evaluation of the likelihood function for the Hydrogen Epoch of Reionization Array (HERA) power spectrum experiment \citep{DeBoer:2017}. For a similar purpose, \cite{Schmit:2017} proposed a different emulation technique based on a neural network. In contrast to these approaches which maintain a connection between curves and the physical parameters which generated them, in this paper, we explore our likelihood analytically by using linear basis functions encoding information from training sets.

Although there has not yet been a detection of the 21-cm spectrum, some recent limits have been placed on both the power spectrum \citep[][see also figure 9 in \cite{DeBoer:2017} for a comprehensive review]{Paciga:2013,Parsons:2014,Dillon:2014,Dillon:2015,Ali:2015,Jacobs:2015} and the global signal \citep{Bowman:2010,Voytek:2014,Bernardi:2016,Singh:2017,Monsalve:2017b}.

Although some have suggested that interferometers may be suited to measure the global signal \citep{Presley:2015, Singh:2015}, most experimental efforts use single antennas. Experiments and mission concepts to measure the global 21-cm signal include the Experiment to Detect the Global EoR Signature \citep[EDGES;][]{Bowman:2010,Monsalve:2017b,Monsalve:2017a}, the Dark Ages Radio Explorer \citep[DARE;][]{Burns:2012,Burns:2017}, the Shaped Antenna measurement of the background RAdio Spectrum \citep[SARAS;][]{Patra:2013,Singh:2017}, the Sonda Cosmol\'ogica de las Islas para la Detecci\'on de Hidr\'ogeno Neutro \citep[SCI-HI;][]{Voytek:2014}, the Zero-spacing Interferometer Measurements of the Background
Radio Spectrum \citep[ZEBRA;][]{Mahesh:2014}, the Large-aperture Experiment to detect the Dark Ages \citep[LEDA;][]{Bernardi:2015,Bernardi:2016,Price:2017}, and the Broadband Instrument for Global HydrOgen ReioNisation Signal \citep[BIGHORNS;][]{Sokolowski:2015}.

The main impediment to all of these experiments is the vast strength of the foregrounds ($\sim 10^3-10^4$ K) relative to the expected strength of the 21-cm signal ($\sim 10-100$ mK). Since the observations must be made at VHF radio frequencies ($\sim 20-200$ MHz), the instrument making the observations will have a very large beam which will corrupt the intrinsically smooth foreground spectrum by imprinting its own spatial/spectral characteristics onto it. Likewise, the radiometer system will leave its own features in the data (some visible even after calibration). As these antenna and receiver features are characteristic of the particular instrument, we use Gaussian beams and do not include calibration errors in the results of this paper.

In order to fit or remove the immense beam-weighted foregrounds observed by global signal experiments, a foreground model must be chosen. In most analyses, this model is a polynomial or a polynomial multiplied by a power law. However, these models can also be used to fit spectral features which look like the expected 21-cm signal, leading to large covariances between the foreground and signal models and, hence, large errors and low result significance. Some have attempted to circumnavigate this by proposing a model which only considers polynomials which have special smoothness properties \citep{Rao:2017}. This model is ``loss resistant'', meaning that when the model is used to fit spectra, it cannot fit out shapes similar to the signal. Use of this model, however, requires all spectral features introduced into the data by the instrument which are not as smooth as the foreground to have been removed completely in some cleaning process. The effectiveness of this technique relies on knowledge of all relevant aspects of the instrument (e.g. antenna and receiver reflection coefficients, receiver noise temperature, etc.); if the knowledge is inaccurate, the cleaning process and the rigidity of the foreground model introduce biases. In contrast to this method, here we present an analysis which, instead of requiring precise knowledge of instrument parameters, requires precise knowledge of the modes in which the instrument parameters can vary (e.g. in frequency)---knowledge more feasibly extracted from simulations and lab measurements. This knowledge allows the errors introduced by imperfect cleaning of the data to be fit consistently.

In this first paper in the series, we propose a method in which we simulate a given experiment multiple times to create a ``training set''. Then, Singular Value Decomposition (SVD)---also commonly known as Principal Component Analysis (PCA) and EigenValue Decomposition (EVD)---is used to extract from this training set linear basis functions describing the systematics. After a similar process is performed using a global signal training set, the signal and systematics basis functions are combined to fit the spectra, yielding an estimate of the 21-cm global signal based solely on the input training sets and the data. This is an extremely fast process that also allows us to extract the signal in a model independent parameter space with a unimodal distribution. The second paper in this series will address how the pipeline takes advantage of initial results such as those found here by facilitating the initialization of an MCMC search to constrain physical parameters (Rapetti et al., in preparation; referred as Paper II hereafter). Paper III (Tauscher et al., in preparation) will then detail how, with a realistic instrument, we can take advantage of instrument-induced polarization to differentiate the power of anisotropic sources like our Galaxy from the power of isotropic sources like the 21-cm global signal \citep[see][]{Nhan:2017}.

Others have proposed using SVD in the context of 21-cm cosmology. \cite{Chang:2010} were the first to suggest that SVD could be used to fit out foregrounds of 21-cm experiments. \cite{Vedantham:2014} and \cite{Switzer:2014} both put forth SVD as a possible method to model systematics in data acquired to measure the 21-cm signal. Unlike those works, we develop systematic modes from a training set based on instrumental degrees of freedom and published maps of Galactic emission, rather than infer modes from the data themselves. Our approach is similar to \cite{Leistedt:2014}, who identified principal contaminant modes in a quasar survey based on known systematic maps.

One important aspect of our method is the choice of where to truncate the modes of both signal and systematics. We perform this choice through a grid search for the global minimum of a given Information Criterion (IC). By running the extraction algorithm with many different inputs and changing the IC which is minimized, we compare the statistical accuracy of the signal extractions resulting from each IC. The IC we consider are Akaike's Information Criterion \citep[AIC;][]{Akaike:1974}, the Bayesian Information Criterion \citep[BIC;][]{Schwarz:1978}, the Bayesian Predictive Information Criterion \citep[BPIC;][]{Ando:2007}, and the Deviance Information Criterion \citep[DIC;][]{Spiegelhalter:2002,Spiegelhalter:2014}. Astrophysics has seen studies of model selection using IC in various contexts \citep[see, e.g.,][]{Porciani:2006,Liddle:2007}. However, they come to differing conclusions of which is the best to use, with some preferring the BIC and others the DIC. Using a statistical measure of the bias in our separation of components of the data (e.g. signal and systematics), we offer a method of choosing which IC to minimize in order to better separate those components.

This work also includes a Python package named \texttt{pylinex} (Linear Extraction in Python), described in Appendix~\ref{sec:pylinex}, implementing the methods described in Section~\ref{sec:methodology}. Since its methods are so general, \texttt{pylinex}, which fits for signals hidden in large systematics and Gaussian noise, is a quick and efficient statistical tool with applications outside of 21-cm cosmology and even outside of astrophysics.

\section{Methodology} \label{sec:methodology}

\subsection{Framework} \label{sec:framework}

Consider a data vector $\by$, which is a combination of $N$ different data components $\by_1,\by_2,\ldots,\by_N$ and Gaussian noise $\bn$,
  \begin{equation}
    \by = \bn + \sum_{k=1}^N\bPsi_k\by_k.  \label{eq:data-vector}
  \end{equation}
  In our case, $\by_1$ would be the global 21-cm signal and $\by_k$ for $k>2$ would include all identified systematic effects (e.g. foregrounds). The matrices $\bPsi_1,\bPsi_2,\ldots,\bPsi_N$---referred to as ``expansion matrices'' here since they allow the data components, $\by_k$, to exist in different (usually smaller) spaces than the full data vector, $\by$---encode the observation strategy used in obtaining the data. For example, if the data vector consists of the concatenation of two spectra which contain the same signal but different systematics, then the relevant expansion matrices are
  \begin{equation}
    \bPsi_{\text{signal}} = \begin{bmatrix} \bI \\ \bI \end{bmatrix}, \  \bPsi_{\text{sys 1}}=\begin{bmatrix} \bI \\ \bzero \end{bmatrix}, \text{ and } \bPsi_{\text{sys 2}}=\begin{bmatrix} \bzero \\ \bI \end{bmatrix}. \label{eq:expansion-matrix-example}
  \end{equation}
\noindent The expansion matrices can also be used to model situation dependent modulation of data components\footnote{To illustrate this, we consider a modification to the example which corresponds to the expansion matrices in Equation~(\ref{eq:expansion-matrix-example}). If the signal is expected to be $1-\kappa$ times as strong in the second spectrum as in the first (e.g. if a fraction $\kappa$ of the source of the signal is blocked in the second spectrum), then the signal expansion matrix is modified to 
  \begin{equation*}
    \bPsi_{\text{signal}}=\begin{bmatrix} \bI \\ (1-\kappa)\bI \end{bmatrix}.
  \end{equation*}} or linear transformations of data\footnote{For instance, setting $\bPsi_k$ equal to the Discrete Fourier Transform (DFT) matrix allows one to model the DFT of $\by_k$.}. Note that if the different components of the data, $\by_k$, have different sizes, then the corresponding expansion matrices, $\bPsi_k$, will have different numbers of columns but the same number of rows.
  
  Our task is to separate a single component of the data (say $\by_1$) from $\by$. To do so, we model $\by$ less noise as a sum similar to Equation~(\ref{eq:data-vector}),
  \begin{equation}
    \bmM(\bx_1,\bx_2,\ldots,\bx_N) = \sum_{k=1}^N\bPsi_k\bF_k\bx_k. \label{eq:model definition}
  \end{equation}
  In this expression, $\bF_k$ is a matrix with basis vectors for fitting $\by_k$ as its columns and $\bx_k$ is a column vector of coefficients modulating those basis vectors. With the definitions
  \begin{equation*}
    \bx=\begin{bmatrix} \bx_1 \\ \bx_2 \\ \vdots \\ \bx_N \end{bmatrix} \ \text{ and } \ \bG=\begin{bmatrix} \bPsi_1\bF_1 & \bPsi_2\bF_2 & \ldots & \bPsi_N\bF_N \end{bmatrix},
  \end{equation*}
  the model can be written $\bmM(\bx)=\bG\bx$. If the model is adequate, the data vector $\by$ is real\footnote{If $\by$ is complex, transpose operations in this paper should be replaced by Hermitian conjugates.}, the true model parameters are $\bx$, and the noise $\bn$ has covariance $\bC$, then, up to a constant, the probability of observing $\by$ is
\begin{equation}
  \mL(\by|\bx) \propto \exp{\left\{ -\frac{1}{2}[\by-\bG\bx]^T\bC^{-1}[\by-\bG\bx]\right\}}. \label{eq:likelihood}
\end{equation}
To form the posterior parameter distribution from this likelihood, we must assume a form for the prior parameter distribution. Here, we use the least informative prior possible\footnote{See Appendix~\ref{sec:priors} for both the assumptions required to include informative priors derived from the training sets, and the equations necessary to do so.}: one which is flat over the entirety of parameter space and any transformation of it\footnote{Note that even though such a prior is referred to as improper, i.e. it cannot be normalized, the posterior distribution is well-behaved.}. This means that the likelihood function is proportional to the posterior distribution, which is therefore Gaussian in $\bx$ with covariance $\bS$ and mean $\bxi$ given by
\begin{subequations} \begin{align}
  \bS &= (\bG^T \bC^{-1} \bG)^{-1}, \label{eq:parameter-covariance} \\
  \bxi &= \bS \bG^T \bC^{-1} \by. \label{eq:parameter-mean}
\end{align} \end{subequations}
The maximum likelihood reconstruction of $\by_k$, $\bgamma_k$, its channel covariance, $\bDelta_k$, and its averaged $1\sigma$ root-mean-square error, $\text{RMS}_k$, are given by
\begin{subequations} \begin{align}
  \bgamma_k &= \bF_k\bxi_k~, \label{eq:channel-mean} \\
  \bDelta_k &= \bF_k\bS_{kk}{\bF_k}^T, \label{eq:channel-covariance} \\
  \text{RMS}_k &= \sqrt{\frac{1}{n_k}\sum_{i=1}^{n_k}(\bDelta_k)_{ii}}\;,
\label{eq:channel-rms}
\end{align} \end{subequations}
where $\bxi_k$ is the portion of $\bxi$ containing parameters modeling $\by_k$, $\bS_{kk}$ is the diagonal block of the covariance matrix corresponding to those parameters, and $n_k$ is the number of data channels in the $\bF_k$ basis. These expressions are also valid without the $k$ subscript, where they represent the same quantities for reconstructions of the full data. When removing the $k$ subscript, $\bF_k$ is replaced by $\bG$.

\subsection{Training sets} \label{sec:training-sets}

The expressions above are valid for any choice of the basis vectors in $\{\bF_k\}$\footnote{The only exception to this statement occurs when two or more of the basis vectors are degenerate with one another.}. However, only sets of basis vectors, $\{\bF_k\}$, which capture the variability of their respective data components, $\{\by_k\}$, will provide meaningful constraints. To automate the intelligent choice of such sets of basis vectors, we assume that, for all $k\in\{1,2,\ldots,N\}$, it is possible to simulate a set of curves $\{\bb^{(k)}_i|i\in\{1,2,\ldots,N^{(k)}_b\}\}$ which reasonably characterizes the most important modes of variation of $\by_k$. These modes are then encoded in $\bF_k$ and used in fitting the full data, $\by$. For this reason, the set of curves is known as a training set. The $k^{\text{th}}$ training set is described by an $n_k\times N_b^{(k)}$ matrix, $\bB_k$, with the curves $\bb^{(k)}$ as its columns. For the rest of Section~\ref{sec:training-sets}, it is assumed that we are speaking about only one of the $N$ different components of the data, so the $k$ subscripts and superscripts are neglected.

We define the basis for the given data component as the matrix $\bF$ which, under the constraint $\bF^T\bK^{-1}\bF=\bI$ (where $\bK^{-1} = \bPsi^T \bC^{-1} \bPsi$), minimizes the Total Weighted Square Error (TWSE) of the fits to all curves in the training set, given by
\begin{subequations} \begin{align}
  \text{TWSE} &= \sum_{i=1}^{N_b} W_i \bb_i^T \bPhi^T \bK^{-1} \bPhi \bb_i \\ &= \Tr(\bW \bB^T \bPhi^T \bK^{-1} \bPhi \bB)\,,
\end{align} \end{subequations}
where $\bW$ is a diagonal matrix of weights $W_i$,\footnote{We generally use $\bW=\bI$.} and $\bPhi=\bI-\bF\bF^T\bK^{-1}$ projects out components of $\bb_i$ in the subspace of $\bF$. The desired basis is given in terms of the weighted SVD of $\bB$, which takes the form $\bB=\bU\bSigma\bV^T$ where $\bSigma$ is diagonal with decreasing non-negative numbers on the diagonal, $\bU^T\bK^{-1}\bU=\bI$, and $\bV^T\bW\bV=\bI$ (see Appendix~\ref{sec:wSVD} for implementation details). Given a number of basis vectors to choose $\eta$, the basis $\bF$ defined above is given by the first $\eta$ columns of $\bU$.

SVD is a reliable way of capturing the modes of variation of a single training set, but if it is performed independently on all components of the data, it may not yield the optimal set of basis vectors for the present purpose, which is separating the different components when they are combined in the same dataset and fit simultaneously. Nevertheless, in lieu of a more sophisticated technique, we perform SVD independently on each training set.

\subsection{Model selection} \label{sec:model-selection}

To select from the models formed by different truncations of the SVD basis sets, we set up a framework within which we test figures of merit, known as information criteria, based on the competition between two terms, the goodness-of-fit term that measures the bias in the fit to the data and the complexity term that penalizes the number of parameters used in that fit. We consider the following information criteria for every truncation under consideration: the DIC \citep{Spiegelhalter:2002,Spiegelhalter:2014}, the BPIC \citep{Ando:2007} and the BIC \citep{Schwarz:1978}. For our likelihood function (Equation~\ref{eq:likelihood}), up to constants independent of the parameters, these are given by
\begin{subequations} \begin{align}
  \text{DIC} &= \bdelta^T\bC^{-1}\bdelta + 2N_p, \label{eq:DIC} \\
  \text{BPIC} &= \bdelta^T\bC^{-1}\bdelta + N_p + 2\ \Tr(\bC^{-1} \bDelta \bC^{-1} \bD), \label{eq:BPIC} \\
  \text{BIC} &= \bdelta^T\bC^{-1}\bdelta + N_p\ln{N_c}. \label{eq:BIC}
\end{align} \label{eq:IC} \end{subequations}
$\!N_p$ is the total number of varying modes across all $N$ sets of basis vectors, $N_c$ is the total number of data channels, $\bDelta=\bG\bS\bG^T$, $\bdelta=\bG\bxi-\by$, and $\bD=[\text{diag}(\bdelta)]^2$. Note that while the goodness-of-fit remains the same across the information criteria, the complexity term varies. The AIC \citep{Akaike:1974} and a variant of the DIC where the complexity term is based on the variance of the log-likelihood \citep[][page~173]{Gelman:2013} were also considered but since our model is linear, they are both equivalent to the DIC in Equation~(\ref{eq:DIC}). When truncating, i.e. selecting between our nested SVD models, we choose the model that minimizes the desired criterion. We investigate which information criterion works best in our analysis in Section~\ref{sec:21-cm_model-selection}.

\section{21-cm global signal application} \label{sec:21-cm_application}

\subsection{Simulated data and training sets}

To illustrate our methods, we propose a simple, simulated experiment to measure the global 21-cm signal using dual-polarization antennas which yield data for all 4 Stokes parameters at frequencies between 40 and 120 MHz. For simplicity, we ignore all systematic effects other than beam-weighted foreground emission, such as human generated Radio Frequency Interference (RFI), refraction, absorption and emission due to Earth's ionosphere, and receiver gain and noise temperature variations. The experiment proposed here is similar to a pair of antennas orbiting above the Lunar farside where the ionospheric effects and RFI need not be addressed \citep{Burns:2017}. An instrument training set corresponding to a realistic antenna and receiver will be included in the analyses of Papers II and III.

The data product of our simulated experiment is a set of 96 brightness temperature spectra. The spectra correspond to 4 Stokes parameters and 6 different rotation angles for 4 different antenna pointing directions. The data vector, $\by$, consists of the concatenation of all of the spectra. The noise level of the data, $\sigma$, is roughly constant across the different Stokes parameters and is related to the total power (Stokes $I$) brightness temperature, $T_b$, through the radiometer equation, $\sigma(\nu) = {T_b(\nu)}/\sqrt{\Delta \nu\ \Delta t}$, with a frequency channel width $\Delta\nu$ of 1 MHz and an integration time $\Delta t$ of 1000 hours (split between the different antenna pointing directions and rotation angles about those directions). The data are split into $N=5$ different components---one for the 21-cm signal and one for the beam-weighted foregrounds (which are correlated across boresight angles and frequency) of each pointing. The signal is the same across all 4 pointings while the foregrounds for each pointing only affect the data from that pointing. The expansion matrices encode this fact.

\begin{figure}
  \centering
  \includegraphics[width=0.5\textwidth]{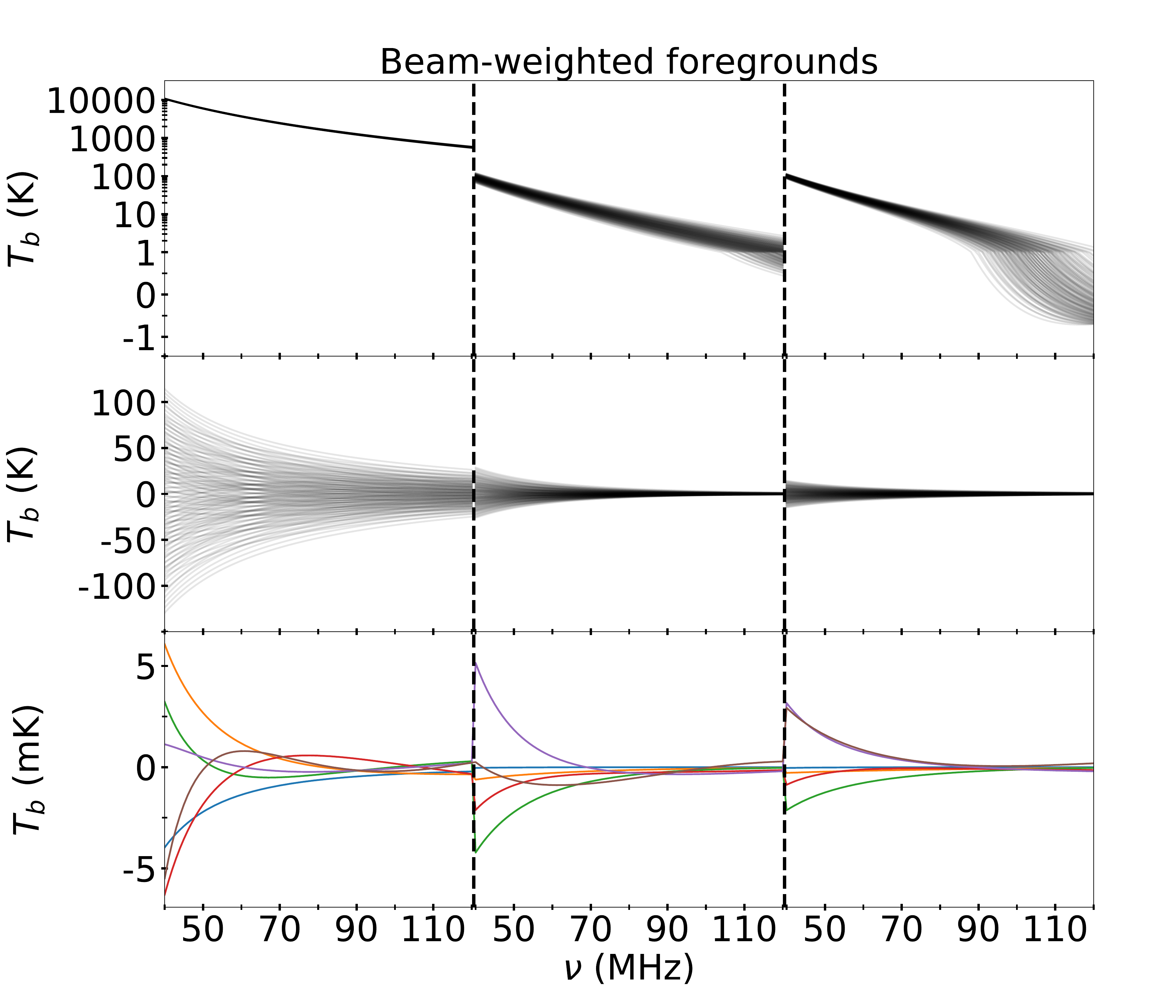}
    \caption{Beam-weighted foreground training set for a single rotation angle about one of the 4 antenna pointing directions (top), the same training set with its mean subtracted (middle), and the first 6 SVD basis functions obtained from the training set (bottom). In the top panel, the y-axis scale is linear between -1 and +1 and logarithmic below -1 and above +1. In addition to the Stokes parameters shown in the three columns (I, Q, U), each curve in the training set contains data for each of the rotation angles $\phi=n\pi/3$ for $n\in\{0,1,...,5\}$. The different rotation angles about the antenna pointing direction are part of the same training set so that SVD can pick up on $\phi$-dependent structure and imprint it onto the basis functions. The modes are ordered from most to least important: blue, orange, green, red, purple, brown. As described in Section~\ref{sec:training-sets}, the modes are normalized so that they yield 1 when divided by the noise level, squared, and summed over frequency, Stokes parameter, and rotation angle about the antenna pointing.} \label{fig:beam_weighted_foreground_training_set_and_basis_functions}
\end{figure}

To create the beam-weighted foreground data for each of the 4 antenna pointing directions, we use the simulation framework of \cite{Nhan:2017}, henceforth referred to as N17. Each Stokes parameter, $\zeta$, observed by the instrument is given by an expression of the form $\int B_{I\rightarrow \zeta}\ T_{\text{gal}}\ d\Omega$ where $T_{\text{gal}}$ is the galaxy brightness temperature and $d\Omega$ is the differential solid angle. As in N17, the 4 relevant beams at frequency $\nu$, polar angle $\theta$, and azimuthal angle $\phi$ are of the form
  \begin{equation}
    \begin{pmatrix} B_{I\rightarrow I}(\nu,\theta,\phi) \\ B_{I\rightarrow Q}(\nu, \theta, \phi) \\ B_{I\rightarrow U}(\nu, \theta, \phi) \\ B_{I\rightarrow V}(\nu, \theta, \phi) \end{pmatrix} \propto b(\nu, \theta)\ \begin{pmatrix} 1 + \cos^2{\theta} \\ -\sin^2{\theta}\ \cos{2\phi} \\ -\sin^2{\theta}\ \sin{2\phi} \\ 0 \end{pmatrix}   \label{eq:beams}
  \end{equation}
  where $b(\nu,\theta) \propto \exp\left[-\theta^2 / 2\alpha^2(\nu)\right]$ and $\alpha(\nu)$ is the angular extent of the beam as a function of frequency. The beam's polarization response converts intensity anisotropy into apparent polarization, while the monopole (cosmological signal) averages to zero in the instrumental Stokes $Q$ and $U$ channels. Measurement in the polarization channels therefore provides discrimination of foreground modes from the signal. Further, rotation about the boresight is used to modulate the polarized components. N17 used this method assuming a spectrally invariant beam, and a sky following a single power law in frequency. Here, with the aid of training sets, we extend the method to allow for spectrally varying beams and an arbitrary sky model. The Galaxy map used in this paper is a spatially dependent power law interpolation between the maps provided by \cite{Haslam:1982} and \cite{Guzman:2011}. The beam-weighted foreground training sets are created using 125 Gaussian beams described by Equation~(\ref{eq:beams}) with varying quadratic models of $\alpha(\nu)$. Figure~\ref{fig:beam_weighted_foreground_training_set_and_basis_functions} shows the training set for one of the 4 antenna pointing directions and some of the corresponding basis functions. Although $V=0$ in this work because $B_{I\rightarrow V}=0$, in real 21-cm global signal experiments with polarimetry, since we expect no circularly polarized light to be incident on the antenna, $V$ can contain useful information about instrument variations.

\begin{figure}
  \centering
  \includegraphics[width=0.5\textwidth]{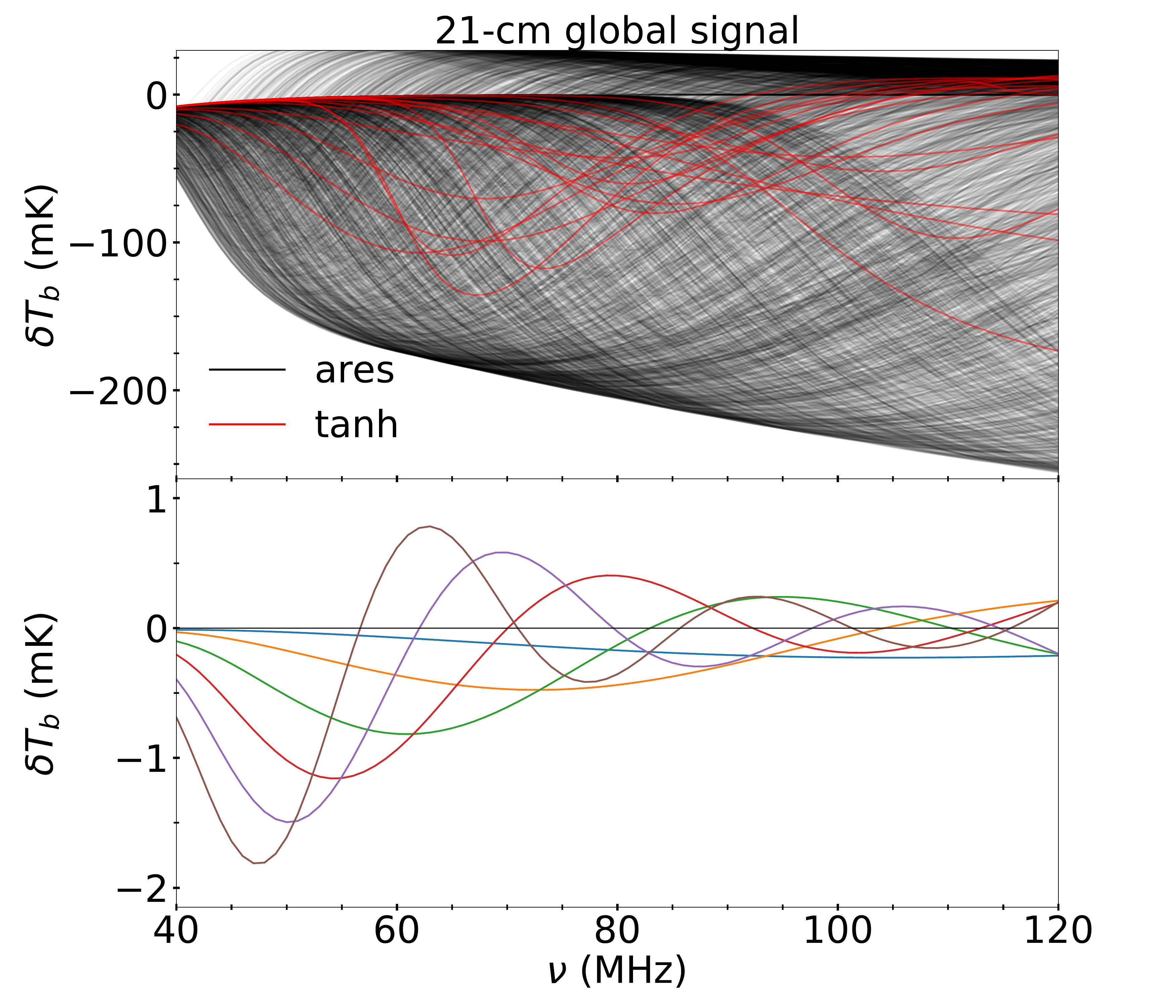}
    \caption{The signal training set used for our analysis was generated by running the \texttt{ares} code $7\times 10^5$ times within reasonable parameter bounds in order to fill the frequency band. The top panel shows a thinned sample of that set (black curves). The red signals are taken instead from the tanh signal model of \cite{Harker:2016}. Such signals will be used in Figures~\ref{fig:signal-bias-statistic-histogram},~\ref{fig:normalized-deviance-histogram}, and~\ref{fig:21-cm-signal-bands}. For illustration purposes, the first 6 SVD basis functions obtained from the \texttt{ares} training set are shown in the bottom panel. The modes are ordered from most to least important: blue, orange, green, red, purple, brown. As described in Section~\ref{sec:training-sets}, the modes are normalized so that they yield 1 when divided by the noise level, squared, and summed over frequency, antenna pointing, and rotation angles about the antenna pointing.}
\label{fig:21-cm_signal_training_set_and_basis_functions}
\end{figure}

The 21-cm global signal training set and a few of the SVD basis functions it provides are shown in Figure~\ref{fig:21-cm_signal_training_set_and_basis_functions}. The training set was created by varying the parameters of the Accelerated Reionization Era Simulations (\texttt{ares}) code\footnote{\url{https://bitbucket.org/mirochaj/ares}}. See \cite{Mirocha:2015,Mirocha:2017a,Mirocha:2017b} for information on the signal models used by \texttt{ares}. In this paper, we ignore the parameter values that make the signals in the training set because we focus only on measuring the brightness temperature profile of the 21-cm signal, not on the astrophysical or cosmological parameters which generated it. Paper II will address inference on these parameters.

\subsection{Optimized truncation} \label{sec:21-cm_model-selection}

\begin{figure}
  \includegraphics[width=0.48\textwidth]{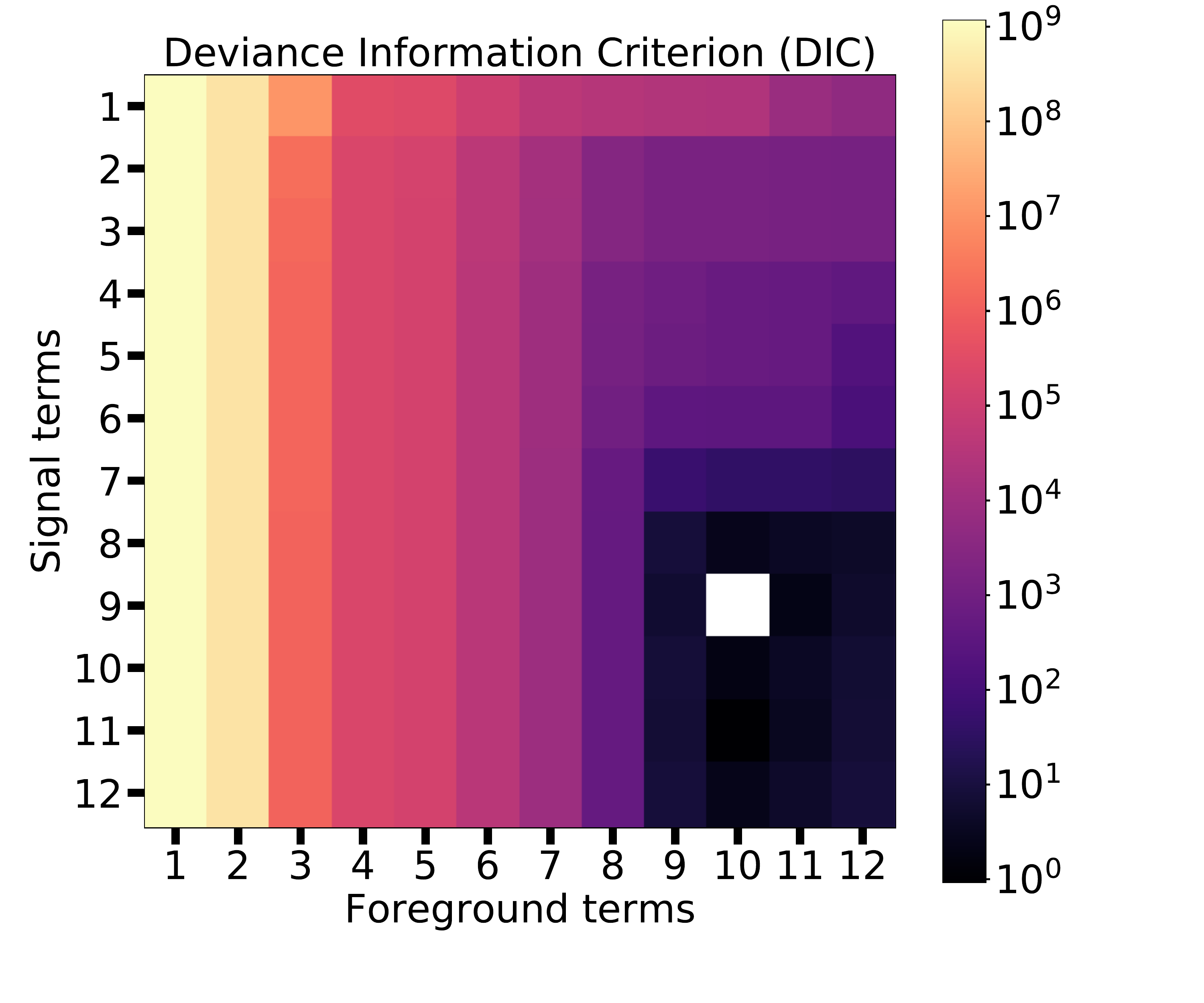}
  \centering
  \caption{Grid of values of the DIC (see Equation~\ref{eq:DIC}) for 144 different fits to simulated data with modes from Figures~\ref{fig:beam_weighted_foreground_training_set_and_basis_functions}~and~\ref{fig:21-cm_signal_training_set_and_basis_functions}. The colors indicate the difference between the DIC and its minimal value. In this case, that minimal value (marked by the white square) occurs when there are 10 foreground and 9 signal modes included in the fit. This same process can be done with any of the information criteria given in Section~\ref{sec:model-selection}. Although only a 12$\times$12 grid is shown here, all of the information criteria were calculated over a 60$\times$30 grid.} \label{fig:DIC-grid}
\end{figure}

The quality of the weighted least squares fit to the data given by the equations in Section~\ref{sec:methodology} depends on the number of parameters describing the signal, $p_{\text{s}}$, and the number of parameters describing the foregrounds of each antenna pointing direction, $p_{\text{f}}$. For each $1\le p_{\text{s}}\le 60$ and $1\le p_{\text{f}}\le 30$,\footnote{As is true here, the maximum number of parameters considered for a given data component should be greater than or equal to the number of modes necessary to fit each curve in the corresponding training set down to the noise level of the data component, which is determined by the covariance matrix $({\bPsi_k}^T\bC^{-1}\bPsi_k)^{-1}$. Recall that $\bC$ is the covariance of the noise, $\bn$, in the data (see Equation~\ref{eq:data-vector}).} we evaluated all of the information criteria in Section~\ref{sec:model-selection}. A section of the $60\times 30$ grid of DIC values for a typical fit where both signal and foregrounds are taken from their training sets is shown in Figure~\ref{fig:DIC-grid}. The model with the lowest DIC is the one with $p_{\text{s}}=9$ and $p_{\text{f}}=10$. See Section~\ref{sec:statistical-tests} for a comparison between the qualities of the fit signals generated by minimizing each of the information criteria.

\subsection{Parameter covariance}   \label{sec:parameter-covariance}

\begin{figure}
  \includegraphics[width=0.5\textwidth]{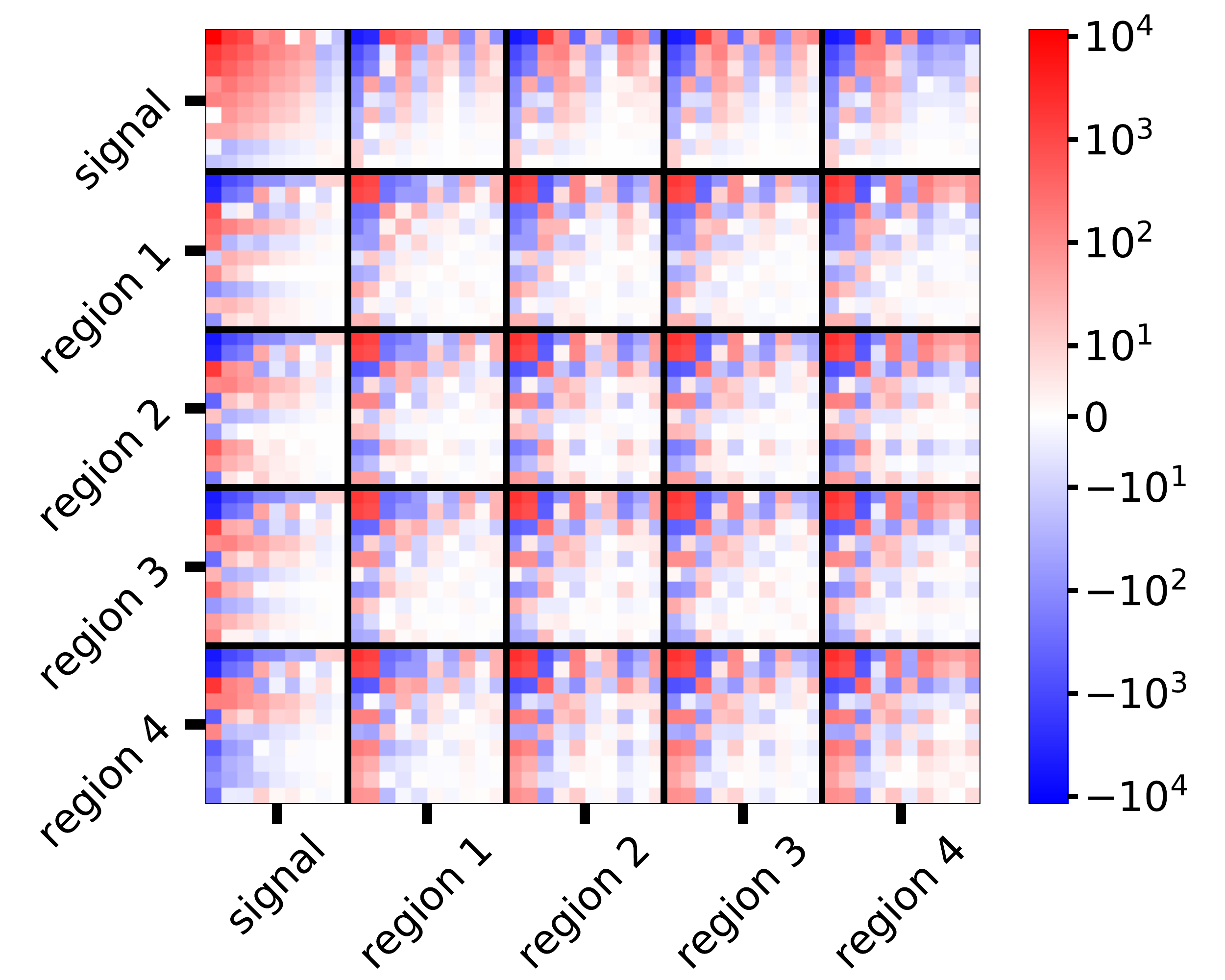}
  \centering
  \caption{Matrix of covariances between the coefficients of all basis functions for the model of the data chosen in the procedure shown in Figure~\ref{fig:DIC-grid} (which has 9 signal modes, and 10 foreground modes for each pointing area). The color scale is linear between $-10$ and $+10$ and logarithmic below $-10$ and above $+10$. Once the numbers of signal and foreground parameters are chosen, these covariances are given by Equation~(\ref{eq:parameter-covariance}). The black horizontal and vertical lines indicate distinctions between different sets of basis functions. As illustrated by Equation~(\ref{eq:channel-covariance}), the error on the final signal estimate depends only on the signal-signal covariance, $\bS_{\text{21-cm}}$ (top left block in this figure).} \label{fig:parameter-covariance}
\end{figure}

The parameter covariance matrix, $\bS$ (Equation~\ref{eq:parameter-covariance}), corresponding to the DIC-chosen model from Figure~\ref{fig:DIC-grid} is shown in Figure~\ref{fig:parameter-covariance}. To aid in intuition about $\bS$, the set of all basis vectors and their overlaps (with respect to the noise covariance, $\bC$) can be viewed as a complete graph---a graph with an edge connecting every pair of vertices---whose vertices belong to $N$ distinct sets. Each vertex is characterized by a pair of numbers $\{i,\alpha\}$ where $i$ is the set of basis vectors the vertex belongs to and $\alpha$ is the specific basis vector from that set. Each edge has a weight, $w_{\{i,\alpha\}\rightarrow\{j,\beta\}}$, given by the overlap between the modes under consideration,
\begin{subequations} \begin{align}
  w_{\{i,\alpha\}\rightarrow\{j,\beta\}} &= w_{\{j,\beta\}\rightarrow\{i,\alpha\}} \\ &= {\boldsymbol{f}}_{\{i,\alpha\}}^T{\bPsi_i}^T\bC^{-1}\bPsi_j{\boldsymbol{f}}_{\{j,\beta\}},
\end{align} \end{subequations}
where $\boldsymbol{f}_{\{i,\alpha\}}$ is the $\alpha^{\text{th}}$ column of $\bF_i$. For the purpose needed here, we define the weight of a path traversing many edges as the product of the weights of its edges, i.e. $w_{\{i,\alpha\}\rightarrow\{j,\beta\}\rightarrow\{k,\gamma\}}=w_{\{i,\alpha\}\rightarrow\{j,\beta\}}\cdot w_{\{j,\beta\}\rightarrow\{k,\gamma\}}$. Assuming the normalization conditions ${\bF_k}^T{\bPsi_k}^T\bC^{-1}\bPsi_k\bF_k=\bI$, the value of the covariance of the coefficients modulating two basis vectors is the sum of the weights of all paths connecting the corresponding vertices of the graph.

From the above, it is clear that the parameter covariances and, hence, errors in the fit are influenced both by the modes included in the basis matrices, $\{\bF_k\}$, and by the expansion matrices, $\{\bPsi_k\}$. The former are controlled by the training sets and the latter are determined by the experimental design. Therefore, errors can be lowered either by modifying the training set to refine the modes used to fit both signal and foreground spectra or by designing an experiment which intrinsically separates the signal and foreground spectra (e.g. through polarization modulation, as employed here).

\subsection{Results} \label{sec:statistical-tests}

\subsubsection{Statistical measures}

To evaluate the behavior of the extraction algorithm, we calculate a few statistical measures for an ensemble of input datasets. In order to assess how accurately the signal is extracted from the data, we first define the signal bias statistic, $\varepsilon$, as
  \begin{equation}
    \varepsilon = \sqrt{\frac{1}{n_{\nu}}\sum_{i=1}^{n_\nu} \frac{[(\bgamma_{\text{21-cm}}-\by_{\text{21-cm}})_i]^2}{(\bDelta_{\text{21-cm}})_{ii}}} \label{eq:signal-bias-statistic}
  \end{equation}
  where $n_\nu$ is the number of frequencies at which the signal is measured, $\by_{\text{21-cm}}$ is the ``true'' (input) signal, and $\bgamma_{\text{21-cm}}$ and $\bDelta_{\text{21-cm}}$ are given by Equations~(\ref{eq:channel-mean}) and~(\ref{eq:channel-covariance}). In an averaged sense, a signal bias statistic of $\varepsilon$ corresponds to a signal bias at the $\varepsilon\sigma$ level. For this reason, if the errors are to be meaningful, the signal bias statistic should be of order unity.

The signal bias statistic is a measure of the accuracy and precision of the signal estimate but it does not gauge the overall quality of the fit to the data. In order to do so, we define the normalized deviance, $D$, through
\begin{equation}
  D = \frac{\bdelta^T\bC^{-1}\bdelta}{N_c-N_p} \label{eq:normalized-deviance}
\end{equation}
where $N_p$ is the number of parameters in the fit, $N_c$ is the number of data channels, and $\bdelta=\bG\bxi-\by$ is the bias in the fit to the data. $D$ is a measure of how well the full set of basis functions (signal and foreground) fits the data. Its true distribution is difficult to ascertain because the number of parameters chosen varies with the inputs; but, since the likelihood is Gaussian and we expect to be able to adequately fit the non-noise component of the data, if $N_p$ parameters are chosen for a given extraction, then the corresponding value of $D$ should follow a $\chi^2$ distribution with $N_c-N_p$ degrees of freedom. Since, unlike $\varepsilon$, $D$ can be calculated even for data whose split between signal and foreground is unknown, it is useful to check its value against confidence intervals generated from the expected distribution to determine if any anomalies were encountered in the extraction process or if the training sets used were inadequate.

\subsubsection{Comparing different inputs and information criteria} \label{sec:information-criteria-comparison}

\begin{figure}
  \centering
  \includegraphics[width=0.48\textwidth]{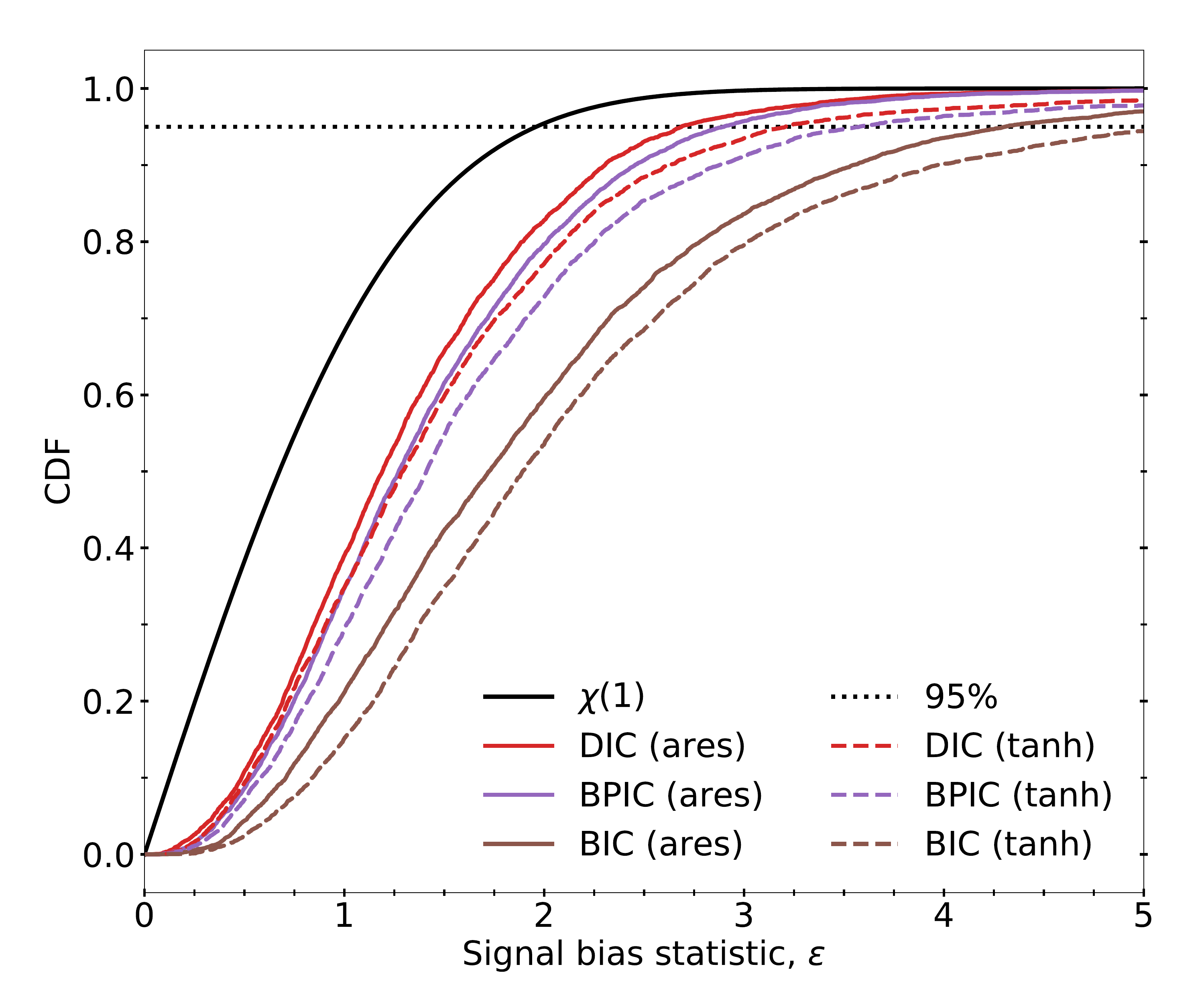}
  \caption{Estimate of the Cumulative Distribution Function (CDF) of the signal bias statistic, $\varepsilon$ (see Equation~\ref{eq:signal-bias-statistic} and text for definition), from 5000 input simulated datasets. The value of the CDF at $\varepsilon=\varepsilon_0$ is equal to the fraction of the 5000 cases where $\varepsilon\le\varepsilon_0$. A bias statistic of $\varepsilon$ roughly corresponds to a bias at the $\varepsilon\sigma$ level. The solid black reference line indicates the CDF associated with a $\chi$ random variable with 1 degree of freedom. This is the distribution which associates 1$\sigma$ with 68\% confidence and 2$\sigma$ with 95\%. To guide the eye, the dotted black line indicates the 95\% level. The realizations which generated the solid lines came from input signals and beam-weighted foreground curves which were both taken from their respective training sets. The dashed lines show the effect of using instead input signals generated using a model (tanh) different than that used for the training set (\texttt{ares}). The color of each curve indicates the statistic which was minimized in order to choose the numbers of parameters of each type (see Sections~\ref{sec:model-selection} and~\ref{sec:21-cm_model-selection}).} \label{fig:signal-bias-statistic-histogram}
\end{figure}

\begin{figure}
  \centering
  \includegraphics[width=0.48\textwidth]{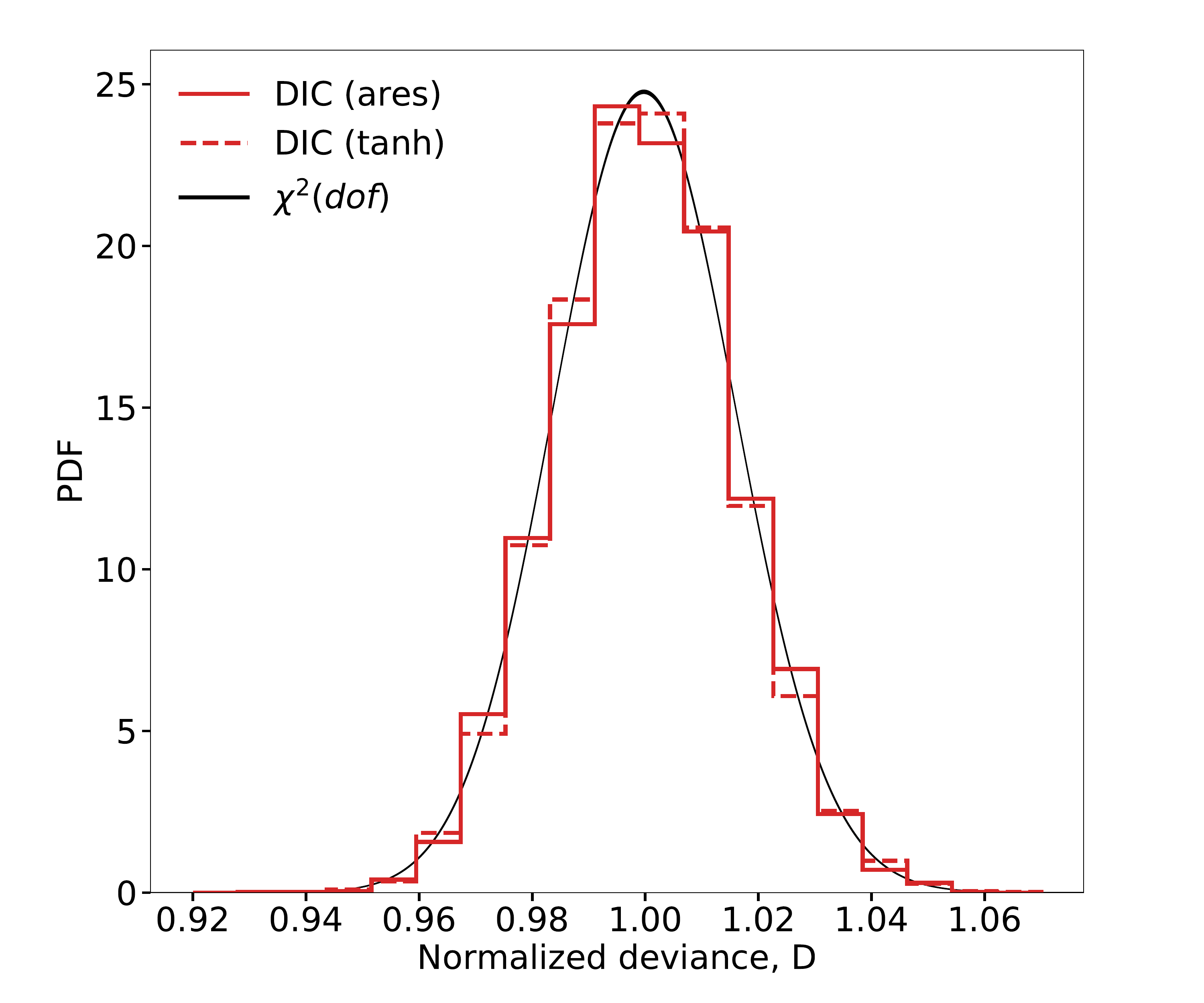}
  \caption{Histogram showing the Probability Distribution Function (PDF) of 5000 values of the normalized deviance, $D$ (see Equation~\ref{eq:normalized-deviance}), for fits with different input signals, beam-weighted foregrounds, and noise when the DIC is used to choose the best model. The solid red line represents the case where the input signal is taken from the signal training set (see Figure~\ref{fig:21-cm_signal_training_set_and_basis_functions}) whereas the dashed red line represents the case where the input signal is generated from a tanh model. For both histograms, the input beam-weighted foregrounds are taken from their training sets. As described in the text, $D$ should follow a distribution approximated by the extremely thin black region, which is a combination of $\chi^2$ distributions associated with the range of degrees of freedom chosen for the extractions. The plot is qualitatively unchanged if we use our other IC for the model selection procedure.} \label{fig:normalized-deviance-histogram}
\end{figure}

We calculated $\varepsilon$ and $D$ for 5000 input datasets, each with their own signal, foregrounds, and noise realization, for two different scenarios. In both cases, the input foregrounds are taken from the training set. In the first case (solid lines in Figures~\ref{fig:signal-bias-statistic-histogram}~and~\ref{fig:normalized-deviance-histogram}), the input signals were taken from the \texttt{ares} training set used to define the modes (shown in Figure~\ref{fig:21-cm_signal_training_set_and_basis_functions}). In the second case (dashed lines in Figures~\ref{fig:signal-bias-statistic-histogram}~and~\ref{fig:normalized-deviance-histogram}), the input signals come from a separate set which parameterizes the ionization history and HI spin temperature as hyperbolic tanh functions of redshift \citep[see][and the red curves in Figure~\ref{fig:21-cm_signal_training_set_and_basis_functions}]{Harker:2016}.

The cumulative distributions of $\varepsilon$ for the two cases described above are shown in Figure~\ref{fig:signal-bias-statistic-histogram}. The different colors represent the different IC from Section~\ref{sec:model-selection} which are minimized in order to choose the number of SVD modes of each type to retain. As expected, when the input signals are not taken from the training set that was used to generate the SVD modes (dashed lines), on average the fits tend to degrade (compare the dashed with the corresponding solid lines). Whether the input signal is in the training set or not, for a given value $\varepsilon_0$ of the statistic, we find that the DIC recovers the global 21-cm signal with $\varepsilon<\varepsilon_0$ more often than any other IC. Thus, we adopt the DIC as our model selection criterion.

Figure~\ref{fig:normalized-deviance-histogram} shows that whether the input signals are taken from the training set or are generated with the tanh model, $D$ follows the expected distribution. Therefore, we find that, combined, the selected SVD signal and foreground modes are adequate to fit the input data down to the noise level.

\subsubsection{Signal estimates} \label{sec:signal-estimates}

\begin{figure*}
  \centering
  \includegraphics[width=0.98\textwidth]{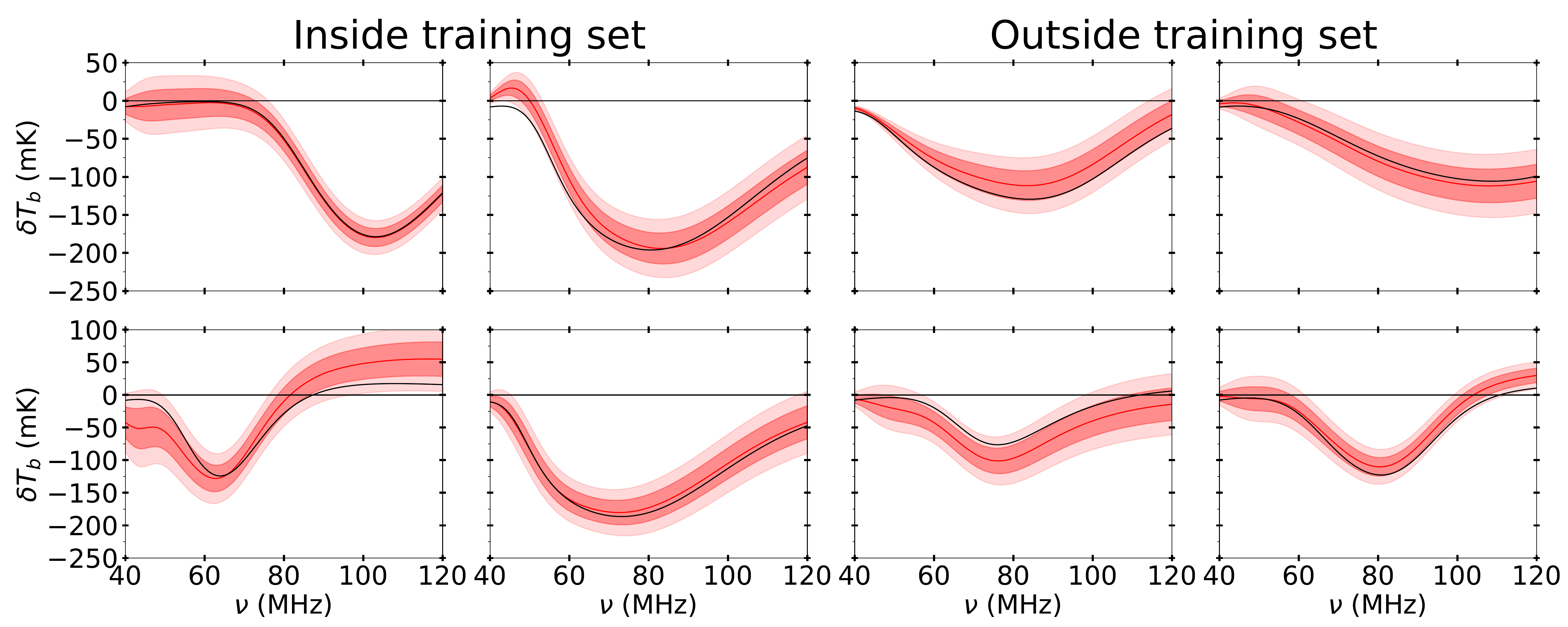}
  \caption{Signal estimates obtained using the linear model defined by SVD eigenmodes calculated from the training sets. The black curves show the input signals, the red curves show the signal estimates, and the dark (light) red bands represent the posterior 1.7$\sigma$ (3.2$\sigma$)---or, equivalently, 68\% (95\%)---confidence intervals (see Figure~\ref{fig:signal-bias-statistic-histogram}~and~Section~\ref{sec:signal-estimates} for details). For all plots, the input beam-weighted foregrounds came from their training sets. The input signals for the four plots on the left came from the signal training set (Figure~\ref{fig:21-cm_signal_training_set_and_basis_functions}) whereas the input signals for the four plots on the right were generated by the tanh model (see Section~\ref{sec:information-criteria-comparison}). Note that by eye there is hardly a difference between these cases, but on average we find that there is, as shown in Figure~\ref{fig:signal-bias-statistic-histogram}.} \label{fig:21-cm-signal-bands}
\end{figure*}

The channel mean and covariance of the estimated global 21-cm signal (in frequency channel space), $\bgamma_{\text{21-cm}}$ and $\bDelta_{\text{21-cm}}$, are given by Equations~(\ref{eq:channel-mean})~and~(\ref{eq:channel-covariance}), respectively. Figure~\ref{fig:21-cm-signal-bands} shows a representative selection of extractions of simulated signals by our analysis. In each panel of the figure, the widths of the bands at frequency $\nu$ are given by 1.7 and 3.2 times the square root of the corresponding diagonal element of $\bDelta_{\text{21-cm}}$, making them 1.7$\sigma$ and 3.2$\sigma$ confidence intervals. Figure~\ref{fig:signal-bias-statistic-histogram} shows that, when the input signal is generated by the tanh model as opposed to being taken from the training set, signals are fit with biases of less than 1.7$\sigma$ (3.2$\sigma$) about 68\% (95\%) of the time. Therefore, the light bands in Figure~\ref{fig:21-cm-signal-bands} show 95\% confidence intervals in the sense that there is a 95\% probability that the squared ratio of the bias (difference between red and black lines) to the error (light red band) averaged across the spectrum is less than 1. Note that this does not mean that there is a 95\% probability that the value of the 21-cm global signal at a given frequency is within the error band. The root mean square widths of the 95\% bands of 95\% of the ares (tanh) input signals are less than 21 (28) mK.

\section{Conclusions} \label{sec:conclusions}

In this paper, we have proposed a method of analysis which transforms the task of extracting a signal from data into one of composing training sets which accurately model the relevant modes of variation in all components of the data. Applying this method to a simple, simulated 21-cm global signal experiment using dual-polarized antennas, we extracted a wide variety of input signals accurately with a 95\% confidence RMS error of $\lesssim 30$ mK. We also showed that, for our purpose of extracting components of data using well separated training sets, minimizing the DIC yields less biased results than minimizing the BIC or BPIC.

For 21-cm experiments analyzed with our method, there are three strategies for decreasing the errors: changing the expansion matrices through modifications of the experimental design (e.g. implementing the polarization technique of N17), changing the basis vectors by using more separable training sets (if possible), and lowering the noise level in the data by adding integration time. The first two of these strategies decrease parameter covariances while the last changes the normalization of the modes, leading to lower errors with the same parameter covariances (see Section~\ref{sec:parameter-covariance}).

As presented in this paper, our method assumes that the posterior distribution on the coefficients of the basis functions is Gaussian. This imposes two requirements: 1) the likelihood function and, hence, all noise must be Gaussian and 2) the different components of the data must be combined by addition. If either of these conditions is violated (e.g. if an expansion matrix, $\bPsi$, is parameter dependent), then the posterior parameter distribution is not Gaussian and therefore must be explored with more sophisticated methods such as MCMC sampling. Such an exploration, however, could still be relatively fast because, although they may be combined in nonlinear ways, the models of the individual components are still linear. Under this variation of the method, model selection as in Section~\ref{sec:model-selection} can still be performed but different expressions for the information criteria must be derived. This more general extraction will be implemented with an MCMC sampler in future versions of \texttt{pylinex}. In addition to generalizing our methods, in future analyses, we will include multiple Galaxy maps in the foreground training set and we will use calibration errors generated by a realistic instrument to generate an instrument training set.

As will be described in Paper II, we have also created a pipeline built around the methods described here to perform inference on physical parameters of the 21-cm signal. After achieving a signal fit like those shown in Figure~\ref{fig:21-cm-signal-bands}, we use it to initialize the iterates of an MCMC sampler tightly packed around or near the likelihood peak. In this MCMC, the 21-cm signal model using SVD basis functions is replaced with one which uses physical parameters. Since there are essentially no constraints on the majority of signal parameters and most realizations of the 21-cm global signal require $\gtrsim 1$ s to compute, such a first guess is important to avoid local minima and is very advantageous towards the convergence of an MCMC chain. 

Therefore, in this paper, we have presented methods to bypass the initial phase of potential traps and the time wasted when an MCMC algorithm wanders through parameter space looking for features in the likelihood by skipping to the stage where the MCMC explores systematically around the likelihood peak. In a simplified yet general setting, we have shown that, with well-characterized training sets and a well-designed experiment, a wide variety of realistic 21-cm global signals can be extracted from simulated sky-averaged observations in the VHF range.

\acknowledgments
We thank Raul Monsalve, Jordan Mirocha, Richard Bradley, Bang Nhan, Licia Verde, and Nicholas Kern for useful discussions. This work was directly supported by the NASA Solar System Exploration Research Virtual Institute cooperative agreement number 80ARC017M0006. This work was also supported by grants from NASA's Ames Research Center (NNA09DB30A, NNX15AD20A, NNX16AF59G) and by a NASA ATP grant (NNX15AK80G). DR is supported by a NASA Postdoctoral Program Senior Fellowship at NASA's Ames Research Center, administered by the Universities Space Research Association under contract with NASA.

\bibliographystyle{aasjournal}
\bibliography{ref}

\appendix
\section{A. Software release}  \label{sec:pylinex}

\texttt{pylinex}\footnote{\url{https://bitbucket.org/ktausch/pylinex}} is a Python package implementing the methods presented in Section~\ref{sec:methodology}. It is split up into 5 modules, the most important of which are the \texttt{expander}, \texttt{basis}, and \texttt{fitter}.

The \texttt{expander} module contains classes which efficiently implement the expansion matrices, $\bPsi$. This is particularly useful when there is a large amount of data involved in the calculations as storing all elements of $\bPsi$ could be unnecessarily memory-intensive, especially when $\bPsi$ is sparse.

The \texttt{basis} module stores classes representing many different types of basis functions, including simple polynomials, Legendre polynomials, Fourier series, and, most importantly for the current work, SVD basis sets.

The \texttt{fitter} module contains the main products of the software: objects which perform fits using the basis sets and expansion matrices from the other modules. The most useful and comprehensive class is the \texttt{Extractor} class which takes as inputs training sets, data with errors, and expansion matrices and yields posterior channel means and errors for each data component as outputs.

\texttt{pylinex} is compatible with Python 2.7+ as well as Python 3.5+ and is implemented efficiently through the use of the \texttt{numpy}, \texttt{scipy}, and \texttt{matplotlib} packages\footnote{\url{https://www.scipy.org/}}. Future versions of \texttt{pylinex} will implement extraction when the model is non-linear and/or the likelihood function is non-Gaussian as described in Section~\ref{sec:conclusions}.

\section{B. Weighted SVD implementation} \label{sec:wSVD}

This section describes our implementation of weighted SVD using only basic SVD algorithms. By weighted SVD, we refer to the decomposition of a real matrix, $\boldsymbol{B}$, of the form $\bB = \bU\bSigma\bV^T$ where $\bSigma$ is diagonal and of the same shape as $\bB$, $\bV^T\bW\bV=\bI$, $\bU^T\bC^{-1}\bU=\bI$, and $\bC$ and $\bW$ are symmetric positive definite matrices. If $\bB$ is a matrix with training set curves as columns, then $\bC$ represents the noise covariance used in fits and $\bW$ is a matrix which weights the different training set curves. To begin, with standard SVD we decompose the matrix $\bC^{-1/2}\bB\bW^{1/2}$ into $\bP\bGamma\bQ^T$ where $\bGamma$ is diagonal, $\bP^T\bP=\bI$ and $\bQ^T\bQ=\bI$. Solving for $\bB$, we find the weighted SVD of $\bB$,
\begin{equation}
  \bB = \bU\bSigma\bV^T = (\bC^{1/2}\bP)\bGamma(\bW^{-1/2}\bQ)^T. \label{eq:wSVD}
\end{equation}
Therefore, the matrices in the weighted decomposition are $\bU=\bC^{1/2}\bP$, $\bSigma = \bGamma$, and $\bV = \bW^{-1/2}\bQ$. In most cases, $\bC$ and $\bW$ are diagonal and the matrix square root is equivalent to the element-wise square root.

Empirically, we find that the weighted SVD algorithm described above scales as $\mO(\text{min}({N_t}^2 N_c, N_t {N_c}^2))$ where $N_t$ is the number of training set curves used and $N_c$ is the number of data channels in each training set curve.

\section{C. Priors from training sets} \label{sec:priors}

Under certain circumstances, it could be beneficial to include prior information on one or more of the data components derived from their training sets. This technique relies on a few key assumptions which limit its applicability. First, the training set must encompass the true multivariate distribution of the observed curves. Second, the prior must be Gaussian for the posterior to be Gaussian, which it must be for simple, analytical expressions to be written for the signal estimate and error. If either of these assumptions is invalid, priors will bias the results.

To generate priors from the training sets, each curve in the training set can be fitted with the SVD modes defined by that training set. Then, the sample mean, $\mu$, and covariance, $\Lambda$, of the points defined by the coefficients of each fit can be found and used to define a multivariate Gaussian prior distribution. Defined in this way, $\mu$ and $\Lambda$ are given by
\begin{equation}
  \bmu = \frac{1}{N_b}\bPi\bZ^T\bJ \ \ , \ \ \bLambda = \frac{1}{N_b-1}\bPi\bZ^T\left(\bI-\frac{1}{N_b}\bJ\bJ^T\right)\bZ\bPi
\end{equation}
where $\bJ$ is a column vector composed of $N_b$ ones. If $\eta$ basis vectors are used (as in Section~\ref{sec:framework}), the matrix $\bPi$ is the top left $\eta\times\eta$ block of $\bSigma$ and the matrix $\bZ$ is the first $\eta$ columns of $\bV$ (where $\bSigma$ and $\bV$ are defined as in Equation~\ref{eq:wSVD}). Including this prior in the fit, the posterior parameter covariance $\bS$ and parameter mean $\bxi$ given by Equations \ref{eq:parameter-covariance} and \ref{eq:parameter-mean} are modified to
\begin{equation}
  \bS = (\bG^T\bC^{-1}\bG + \bLambda^{-1})^{-1} \ \ \text{ , } \ \ \bxi = \bS(\bG^T\bC^{-1}\by + \bLambda^{-1}\bmu).
\end{equation}

\end{document}